\documentclass[prl,a4paper,twocolumn,superscriptaddress]{revtex4-1}

\usepackage[utf8x]{inputenc}
\usepackage{amssymb}
\usepackage{amsmath}
\usepackage{graphicx}
\usepackage{psfrag}
\usepackage{latexsym}
\usepackage{hyperref}

\newcommand{\del}{\partial}
\newcommand{\calv}{{\cal V}}
\newcommand{\calo}{{\cal O}}
\newcommand{\call}{{\cal L}}
\newcommand{\dd}{{\delta}}

\newcommand{\eq}{\begin{equation}}
\newcommand{\eqx}{\end{equation}}
\newcommand{\eqn}{\begin{eqnarray}}
\newcommand{\eqnx}{\end{eqnarray}}

\newcommand{\reef}[1]{(\ref{#1})}

\newcommand{\eg}{{\it e.g.,}\ }
\newcommand{\ie}{{\it i.e.,}\ }

\begin{document}

\title{Equilibration rates in a strongly coupled nonconformal quark-gluon plasma} 

\author{Alex Buchel}\email{abuchel@perimeterinstitute.ca}

\affiliation{\it Department of Applied Mathematics, Department of Physics and Astronomy,
University of Western Ontario, London, Ontario N6A 5B7, Canada}

\affiliation{\it Perimeter Institute for Theoretical Physics, Waterloo, Ontario N2L 2Y5, Canada} 

\author{Michal P. Heller}\email{mheller@perimeterinstitute.ca}

\altaffiliation[On leave from: ]{\emph{National Centre for Nuclear Research,  Ho{\.z}a 69, 00-681 Warsaw, Poland.}}

\affiliation{\it Perimeter Institute for Theoretical Physics, Waterloo, Ontario N2L 2Y5, Canada} 

\author{Robert C. Myers}\email{rmyers@perimeterinstitute.ca}

\affiliation{\it Perimeter Institute for Theoretical Physics, Waterloo, Ontario N2L 2Y5, Canada}

\begin{abstract}

We initiate the study of equilibration rates of strongly coupled quark-gluon plasmas in the absence of conformal symmetry. We primarily consider a supersymmetric mass deformation within \mbox{${\cal N}=2^{*}$}~gauge theory and use holography to compute quasinormal modes of a variety of scalar operators, as well as the energy-momentum tensor. In each case, the lowest quasinormal frequency, which provides an approximate upper bound on the thermalization time, is proportional to temperature, up to a pre-factor with only a mild temperature dependence. We find similar behaviour in other holographic plasmas, where the model contains an additional scale beyond the temperature. Hence, our study suggests that the thermalization time is generically set by the temperature, irrespective of any other scales, in strongly coupled gauge theories.

\end{abstract}

\maketitle

\emph{Introduction and summary.--} The discovery of the quark-gluon plasma (QGP) in ultrarelativistic heavy ion collisions \cite{Adams:2005dq,Aamodt:2010pa} is one of the landmark breakthroughs of the 21$^{\rm st}$ century physics. Its behaviour as a strongly-coupled liquid made it clear that our ability to describe these experiments in an ab initio manner directly from quantum chromodynamics (QCD) is insufficient. In particular, one outstanding theoretical puzzle is elucidating microscopic mechanisms behind the rapid appearance of the QGP itself (with temperature of order of 500 MeV \cite{Broniowski:2008vp}) in less than 1 fm~\cite{Heinz:2001xi}.

The absence of reliable theoretical tools to describe time-dependent phenomena in QCD away from the regime of small coupling brought a great deal of attention to strongly-coupled gauge theories solvable using holography (\eg see \cite{Chesler:2015lsa}, for a recent review). In holography, the QGP creation is associated with the formation of a black hole \cite{Chesler:2008hg}, and the QGP equilibration with the relaxation of the dual black hole  \cite{Horowitz:1999jd,Son:2002sd}. In the latter, equilibration rates are associated with imaginary parts of the dual black hole quasinormal modes (QNM) frequencies. 

Extensive studies of the simplest holographic models in which gauge theories are conformally invariant, most notably of ${\cal N}$ = 4 super Yang-Mills theory (SYM), found near- and far-from-equilibrium relaxation times not greater than the inverse of the final temperature (\eg see \cite{Kovtun:2005ev,Chesler:2008hg}). As the phenomenological estimates obey \mbox{1~fm $\times$ 500 MeV = ${\cal O}$(1)}, one may conclude that the large value of the QCD coupling constant is the factor responsible for fast equilibration of the QGP. However, this conclusion is premature, as, at least near equilibrium, the temperature provides the only scale in the ${\cal N}$ = 4 SYM plasma and dimensional analysis dictates that the equilibration rate is proportional to the temperature.

\begin{figure}
\includegraphics[width=0.47\textwidth]{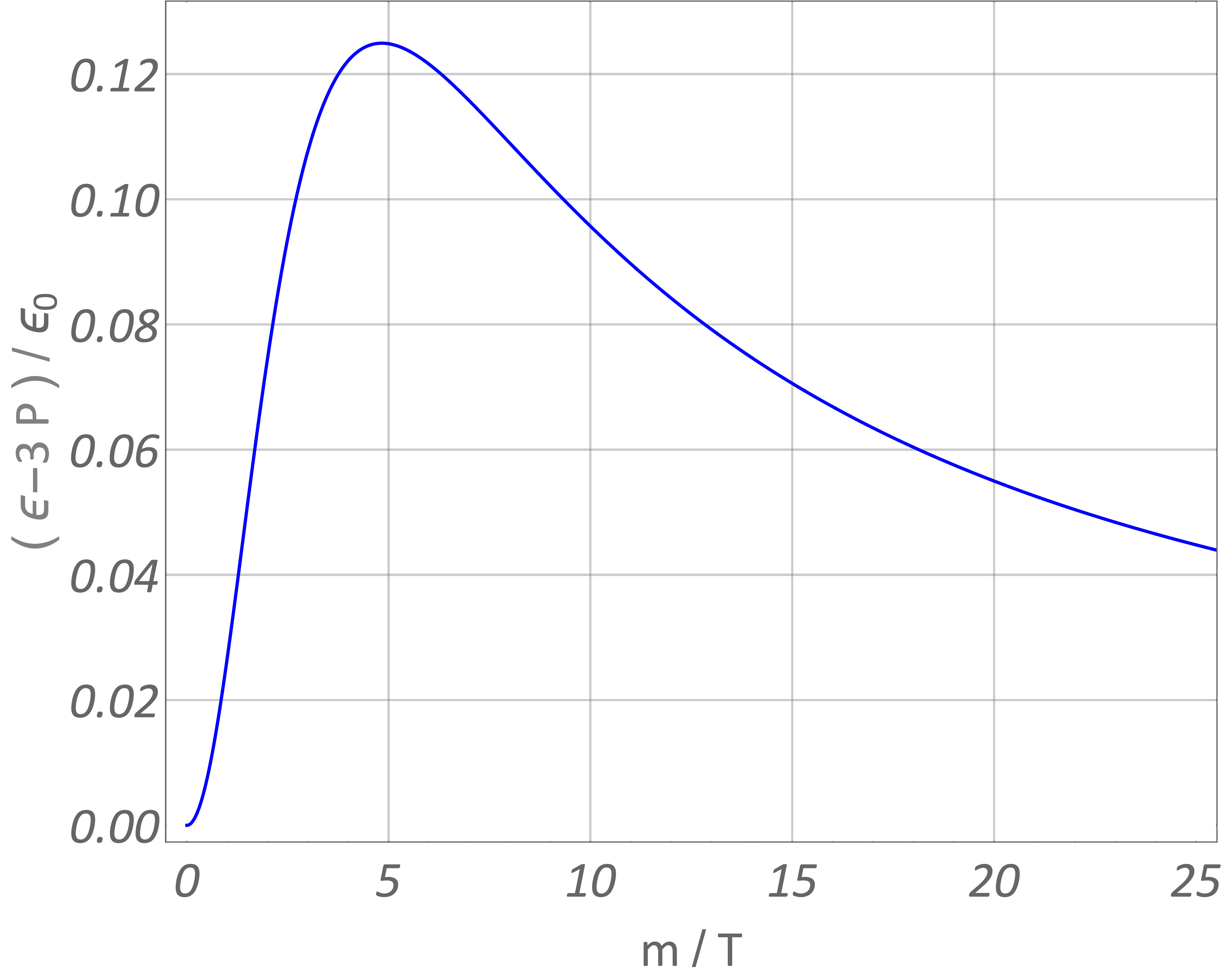}
\caption{Trace of the energy-momentum tensor normalized to the energy density of ${\cal N} = 4$ SYM ($\epsilon_{0} = \frac{3}{8} \pi^{2} N_{c}^{2} T^{4}$ with $N_{c}$ denoting the number of colours) as a function of $m/T$. The results indicate that, thermodynamically, the effects of the conformal symmetry breaking are the strongest at $m/T \approx 4.8$.}
\label{fig.1}
\end{figure}

Motivated by this crucial point, our letter considers equilibration processes in the holographic model describing the {\it nonconformal} ${\cal N} = 2^{*}$ gauge theory. This theory is obtained as a (supersymmetric) deformation of ${\cal N} = 4$ SYM,  giving mass $m$ to half of its fields \cite{Buchel:2007vy}. The QGP plasma of ${\cal N} = 2^{*}$ gauge theory is characterized by the temperature $T$ and one dimensionless parameter $m/T$, the ratio of  the mass to the temperature. We vary the latter from $m/T = 0$ (\ie the ${\cal N} = 4$ SYM plasma) up to $m/T = 25$. The biggest deviation from conformality occurs at $m/T \approx 4.8$ (see~Fig.~\ref{fig.1}) \footnote{At $m/T\approx 4.8$, $(\epsilon-3P)/\epsilon \approx 20\%$ in the ${\cal N}=2^*$ plasma while lattice estimates for the analogous quantity in QCD are somewhat higher, \ie approximately $50\%$ where $(\epsilon-3P)/T^4$ peaks \cite{Bazavov:2014pvz}.}. 

\begin{figure}
\includegraphics[width=0.47\textwidth]{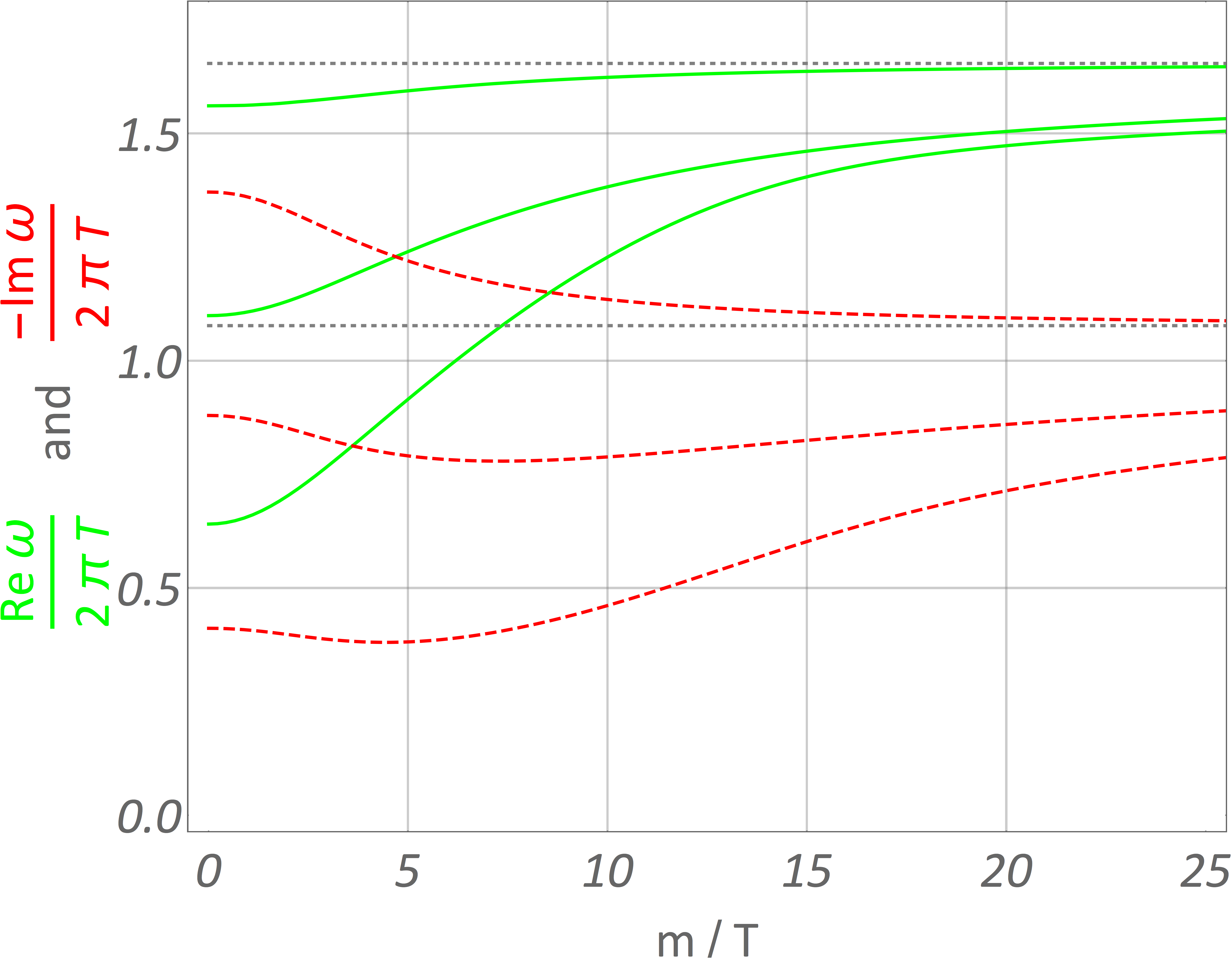}
\caption{Real (green continuous) and minus imaginary (red dashed) parts of the lowest quasinormal mode frequencies for phenomenological scalar operators of dimensions $\Delta = 2, 3$ and $4$ (from bottom to top). The frequencies do not change significantly as a function of m/T, which leads to universal equilibration in 1/T. One can also infer from this plot that all the frequencies asymptote at low temperatures to the quasinormal mode of a massless scalar field living in the (1+5)-dimensional AdS-Schwarzschild geometry (dotted curves). For explanation, see the supplemental material.}  
\label{fig.2}
\end{figure}

We primarily focus on the behaviour of small homogeneous perturbations of ${\cal N} = 2^{*}$ plasma given by phenomenological operators of dimensions $2$, $3$ and $4$, top-down operators of dimensions 2 and 3 and the energy-momentum tensor \footnote{Such perturbations, even though they carry no momentum, are nevertheless responsible for relaxation towards equilibrium in holographic models of expanding plasma systems \cite{Janik:2006gp,Heller:2013fn}}. We compute the frequencies of the corresponding least-dampened QNM, as they set an approximate upper bound on thermalization times. We observe that QNM frequencies, when normalized to temperature, change rather mildly as a function of $m/T$ (see Fig. \ref{fig.2}, \ref{fig.3} and \ref{fig.4}). The largest change in frequencies with respect to ${\cal N}=4$ SYM is roughly a factor of two and is typically much less. This implies that the equilibration time for generic small perturbations is still set by $T$, as in the conformal case \cite{Kovtun:2005ev}.

To test the robustness of this finding, we also considered other nonconformal holographic models. While we do not present the details, 
examining the (nonsupersymmetric) bosonic mass deformation of ${\cal N}=4$ SYM~\cite{Buchel:2007vy} yields qualitatively similar behaviour for all of the above quantities. We also consider QNM of a strongly coupled conformal plasma with chemical potential $\mu$. This constitutes another example of equilibrium state specified by a single dimensionless number, \ie $\mu/T$. We analyze the behaviour of phenomenological operators of dimensions $2$, $3$ and $4$. We find that there is a regime in which the lowest QNM frequencies, when normalized to temperature, change mildly as a function of $\mu/T$ (see Fig.~\ref{fig.RN}).

\begin{figure}
\includegraphics[width=0.47\textwidth]{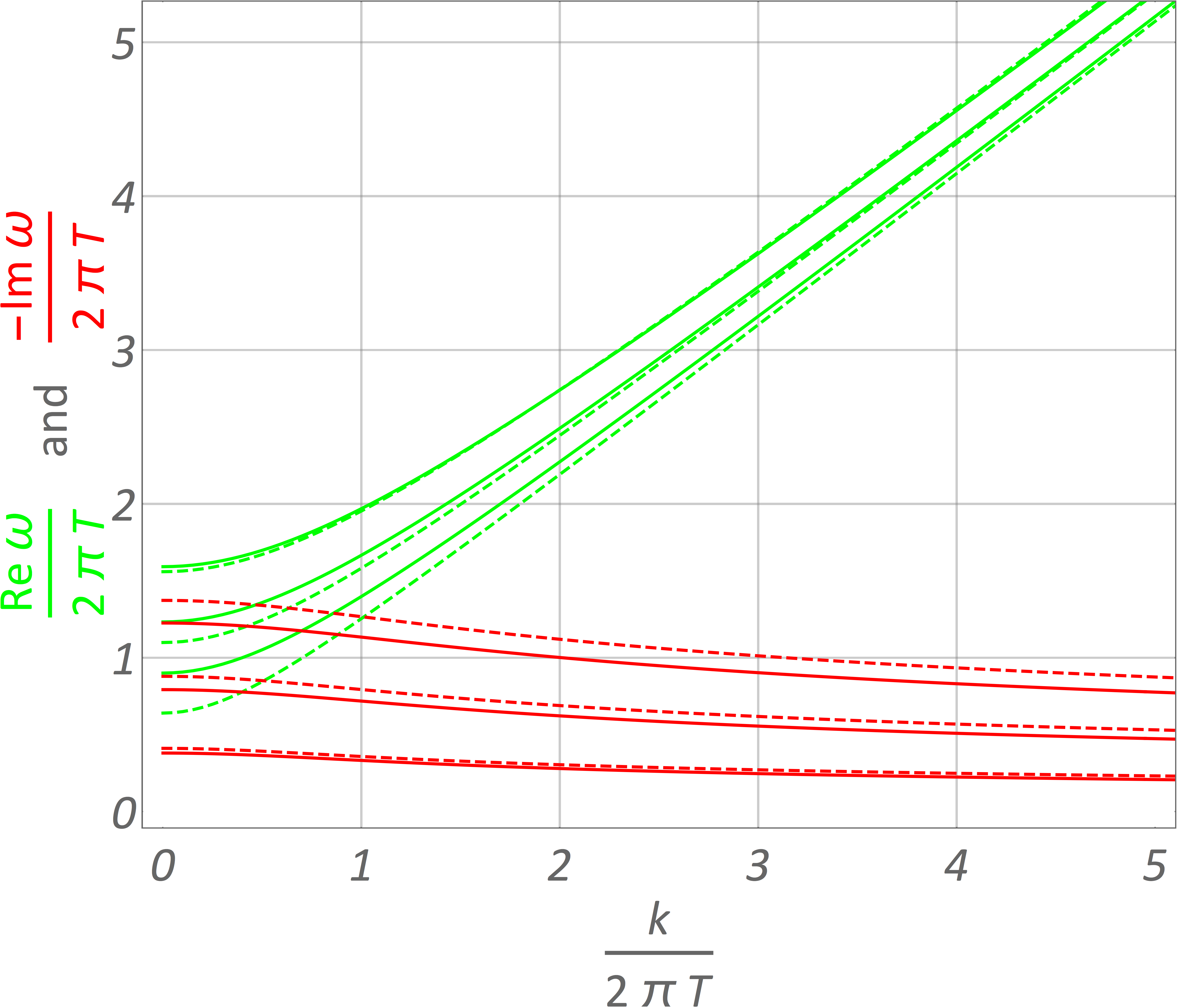}
\caption{Momentum dependence of the real (green) and minus imaginary part (red) of the QNM frequency of phenomenological operators with $\Delta = 2, 3$ and $4$ (from bottom to top) for $m/T = 0$ (${\cal N} = 4$ SYM, dashed) and $m/T = 4.8$ (continuous). Surprisingly, corresponding curves are very close to each other despite of the fact that $m/T = 4.8$ matches the locus of the maximal deviation from conformal invariance in thermodynamics of ${\cal N} = 2^{*}$ (see Fig. \ref{fig.1}).}  
\label{fig.k}
\end{figure}

\begin{figure}
\includegraphics[width=0.47\textwidth]{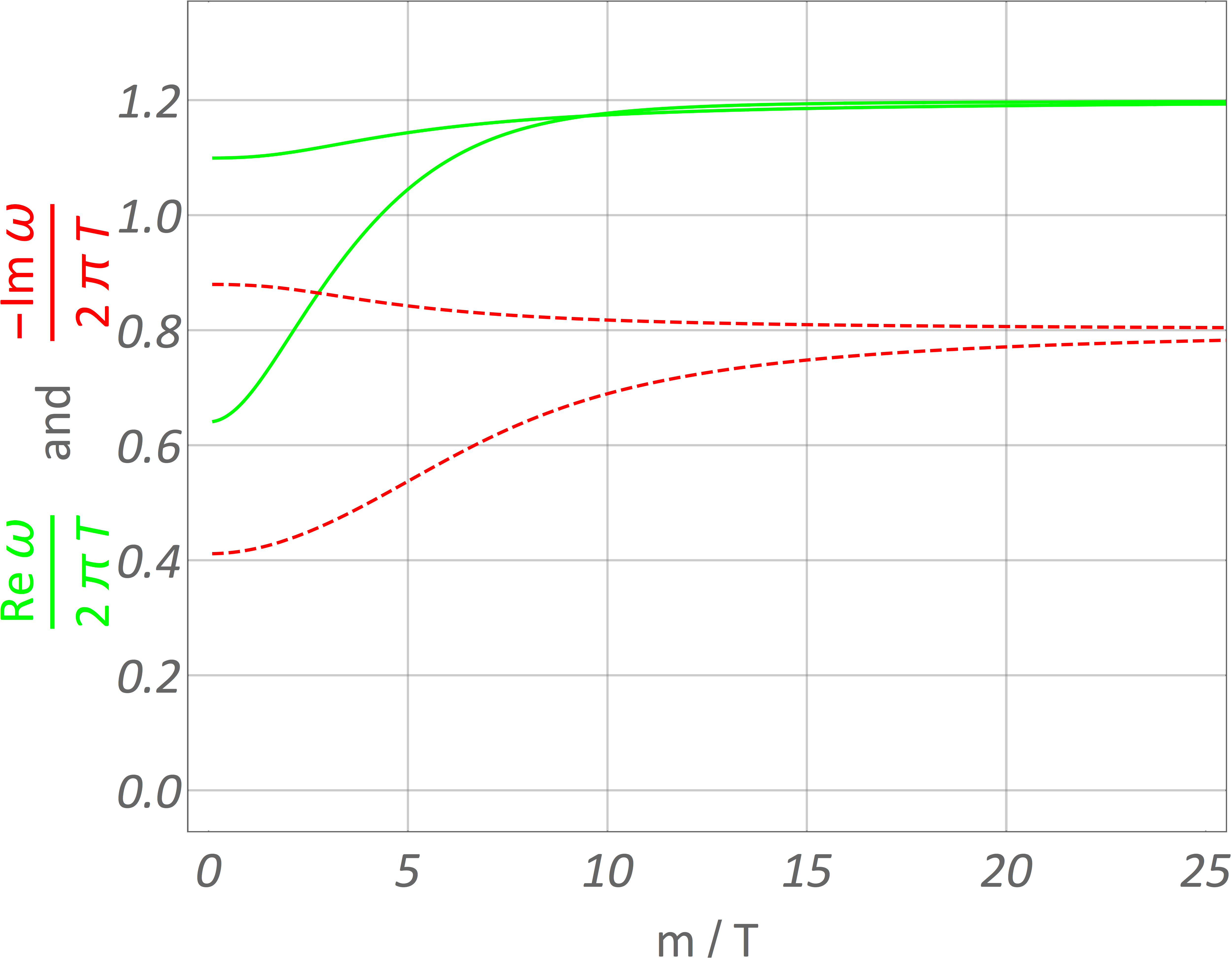}
\caption{Real (green continuous) and minus imaginary (red dashed)
  parts of the lowest quasinormal mode frequencies for
 $\calo_3$  and $\calo_2$ operators (see \eqref{linear})  
in ${\cal N}=2^*$ gauge theory (from top to bottom). The frequencies
  do not change significantly as a function of m/T.
}
\label{fig.3}
\end{figure}

\begin{figure}
\includegraphics[width=0.47\textwidth]{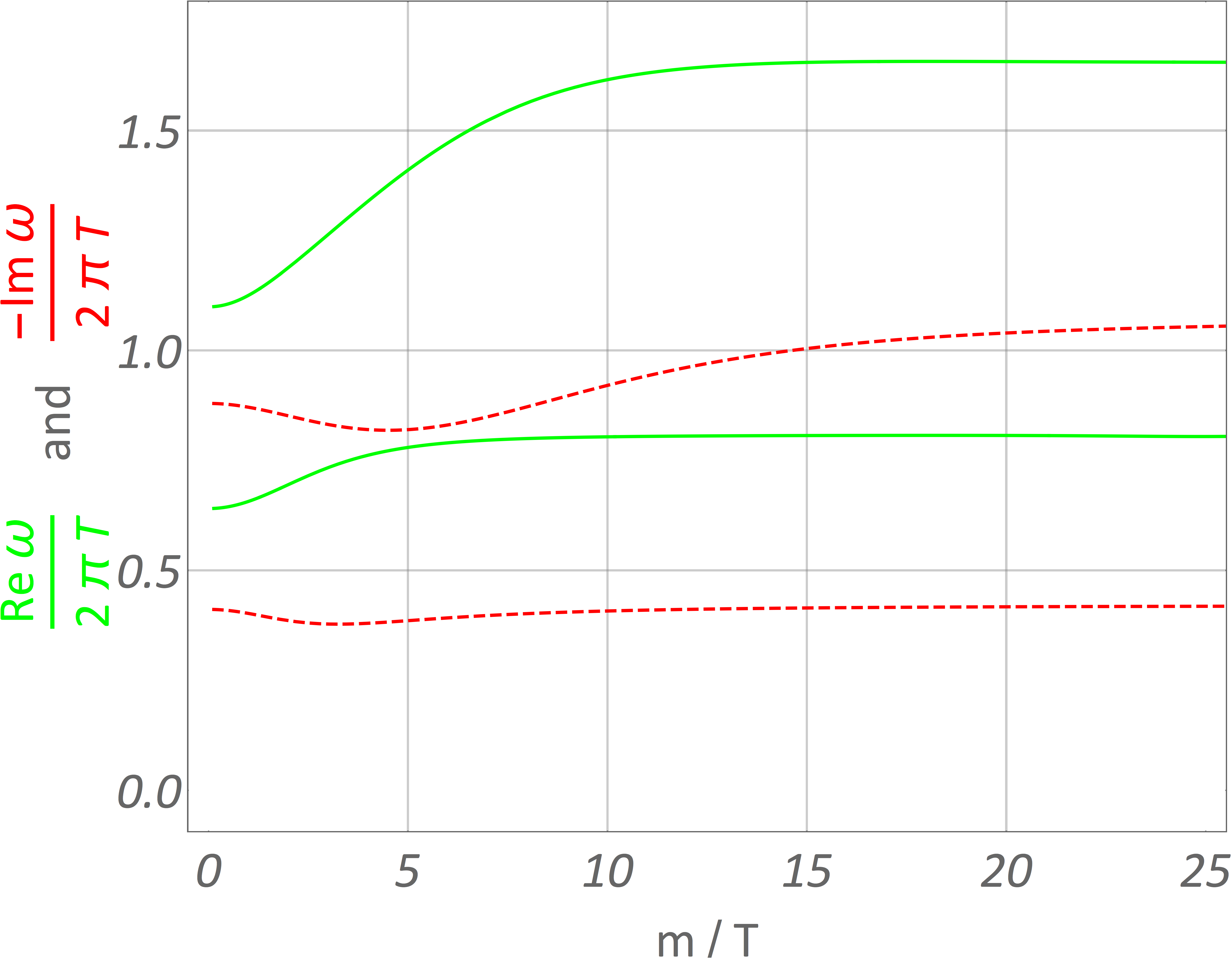}
\caption{Real (green continuous) and minus imaginary (red dashed)
  parts of the lowest quasinormal mode frequencies 
  the operators $\calo_\chi$ and $\calo_\alpha$ 
  (from top to bottom) in ${\cal N}=2^*$ gauge theory. 
}
\label{fig.4}
\end{figure}

\begin{figure}
\includegraphics[width=0.47\textwidth]{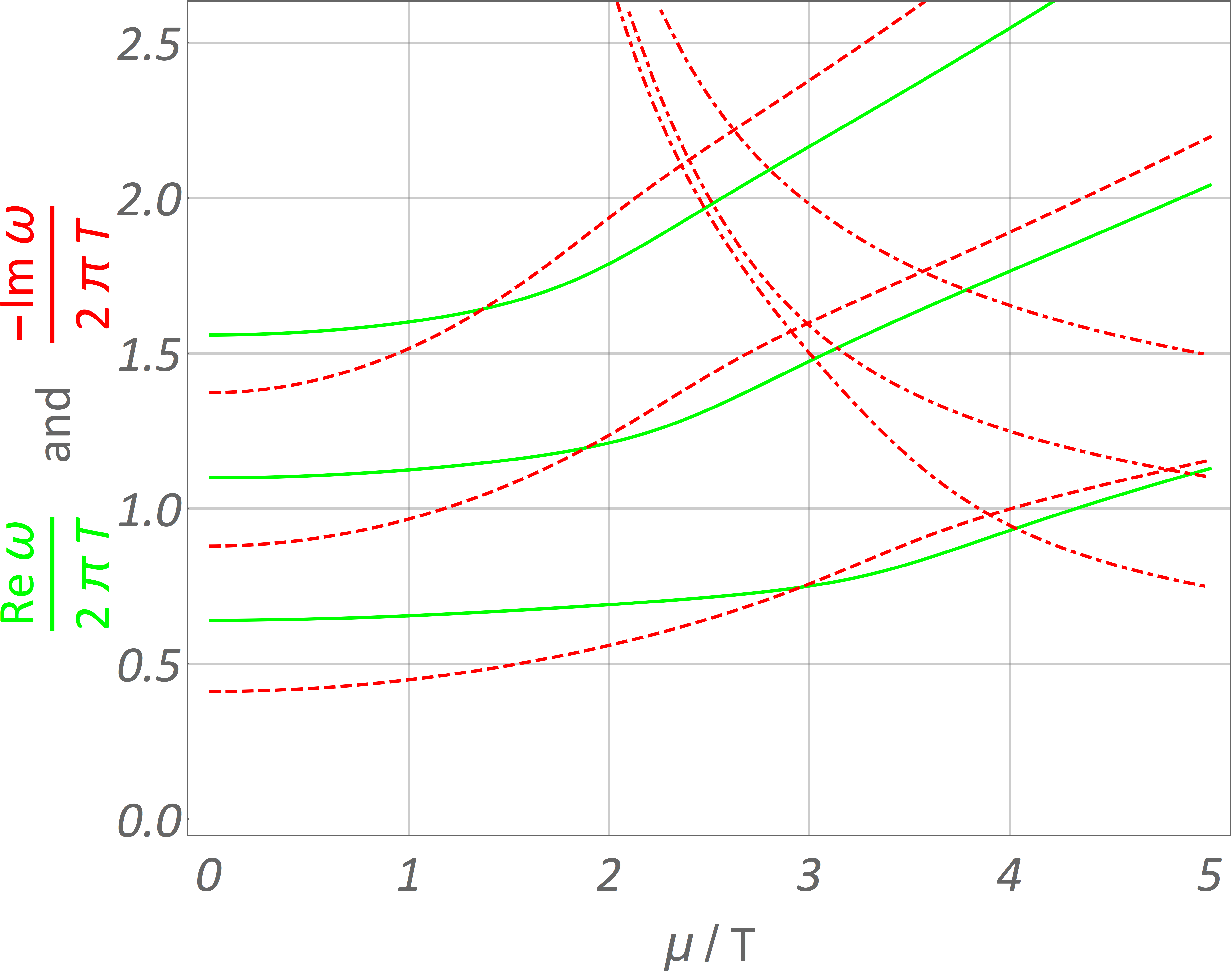}
\caption{Real (green continuous) and minus imaginary (red dashed) parts of the QNM frequencies originating at $\mu/T = 0$ from the lowest QNM for phenomenological scalar operators of dimensions $\Delta = 2, 3$ and $4$ (from bottom to top) in the AdS-RN background. Red dot-dashed curves denote purely imaginary QNM for operators of dimensions $\Delta = 2, 3$ and $4$ (also from bottom to top), which are special to finite $\mu$ and for sufficiently large $\mu/T$ become the least-dampened modes (see \cite{Edalati:2010hk,Bhaseen:2012gg} for the discussion of this branch). Before this transition occurs, the QNM frequencies change by a small amount, similar to what we see for ${\cal N} = 2^{*}$ gauge theory. Despite the transition, the equilibration time remains to be bounded by 1/T, in accord with the main message of this letter. Finally, note that it is not possible to relate in any meaningful way the values of parameter $\mu/T$ appearing here and $m/T$ of ${\cal N} = 2^{*}$ gauge theory, as they describe two different physical systems.}
\label{fig.RN}
\end{figure}

Finally, let us notice that there is strong evidence that equilibration in $1/T$ also occurs for large deviations from equilibrium ~\cite{Buchel:2012gw,Buchel:2014gta,Heller:2012km,Heller:2013oxa,FuiniYaffe}. Here, our results in Fig.~\ref{fig.k} show that including momentum dependence does not change the qualitative behaviour of ${\cal N}=2^{*}$ QNM, as compared to ${\cal N}=4$ SYM. Moreover, the so-called close limit approximation \cite{Price:1994pm,Anninos:1995vf}, originally devised for black hole mergers in asymptotically flat spacetimes, suggests that nonlinear effects as seen by the expectation values of local operators in a boundary field theory are not decisive once a single horizon forms in the bulk \cite{Heller:2012km,Heller:2013oxa,FuiniYaffe}. We are thus led to conjecture that equilibration in \emph{any} strongly coupled gauge theory, once a single horizon forms in a dual gravity description, will be captured in a satisfactory way by ${\cal N} = 4$ SYM (see \cite{Cardoso:2013vpa,Craps:2013iaa} for evolution of bottom-up nonconformal holographic models before this occurs). For holographic descriptions of heavy ion collisions, this suggests the most fruitful direction will be relaxing geometric symmetry constraints in the dynamics in ${\cal N} = 4$ SYM (\eg see \cite{Chesler:2015wra,Bantilan:2014sra}) rather than writing complicated codes to incorporate broken conformal symmetry.

\emph{${\cal N}=2^{*}$ thermodynamics.--} The holographic description of ${\cal N}=2^{*}$ QGP is a five-dimensional black hole
\eq
\label{eq.metric}
ds_{5}^{2} = e^{2 A(z)} \left( - B(z)^{2} dt^2 + d\mathrm{\textbf{x}}^{2}\right) + L^{2} \frac{dz^{2}}{z^2},\eqx
where $L$ sets the overall curvature scale and the boundary is at $z = 0$. The metric \eqref{eq.metric} solves the equations of motion derived from the following effective action
\eq
\label{eq.Lgrav}
S = \frac{1}{16 \pi G_{5}} \int \mathrm{d}^{5} x \sqrt{- g} \left(R + {\cal L}_{\mathrm{matter}} \right),
\eqx
where the matter part, as discovered by Pilch and Warner in \cite{Pilch:2000ue} (PW), is given by
\eq
\label{eq.Lmatter}
- {\cal L}_{\mathrm{matter}} = - {\cal L}_{PW} = 12 (\partial \alpha)^{2} + 4 (\partial \chi)^{2} + V 
\eqx
and the scalar potential takes the form
\eq
\label{eq.VPW}
V =  \frac{1}{L^{2}} \left[ - 4 e^{-4 \alpha} - 8 e^{2 \alpha} \cosh{2 \chi} + e^{8 \alpha} \sinh^{2}{2 \chi} \right].
\eqx
The mass $m$ in the ${\cal N} = 2^{*}$ theory 
is encoded in the near-boundary behaviour of the scalars as
\eq
\label{eq.scalarbdry}
\chi = \frac{m L}{2} z +{\cal O}(z^{2}\log{z}), \quad \alpha = \frac{(m L)^{2}}{6} z^{2} \log{z} + {\cal O}(z^{2}).
\eqx
We choose the units in which the event horizon is located at $z = 1$  (\ie $B(1) = 0$). The QGP temperature is then given by $T =e^{A(1)} B'(1)/(2\pi L)$ and there exists a one-parameter family of geometries labeled by the ratio $m/T$. We construct those geometries using a shooting method. See  \cite{Pilch:2000ue,Buchel:2000cn,Buchel:2003ah,Buchel:2004hw,Buchel:2007vy,Hoyos:2011uh} for details of ${\cal N}=2^*$ holography, its equilibrium thermodynamics and numerical implementation of the shooting method.

As a benchmark of ${\cal N} = 2^{*}$ thermodynamic properties, in Fig. \ref{fig.1} we consider the trace of the stress tensor normalized to the energy density of ${\cal N} = 4$ SYM and plot this ratio as a function of $m/T$. This quantity starts at $0$ for $m/T = 0$, as ${\cal N} = 4$ SYM is conformally invariant, and peaks at $m/T \approx 4.8$. We thus expect the strongest violations of conformality to occur precisely there. 

\vspace{5 pt}

\emph{Thermalization time for phenomenological scalar operators.--} The simplest equilibration process to consider is perturbing a holographic QGP by exciting a phenomenological operator $O$ of dimension $\Delta$. The gravity description of the latter is a minimally coupled scalar $\psi$:
\eq
\label{eq.Sscalar}
S = \frac{1}{4 \pi G_{5}} \int \mathrm{d}^{5} x \sqrt{-g} \left((\partial \psi)^{2} - M^{2} \psi^{2} \right)
\eqx
where the mass is given by $M^2=\Delta(\Delta-4)/L^2$. 

We search for solutions with a harmonic dependence on $t$ and $x$, \ie $\psi = e^{- i \omega t + i k x} f(z)$. We use the shooting method to search for the normalizable modes satisfying an ingoing boundary condition at the horizon. These conditions are satisfied by a discrete set of complex frequencies that define QNM. In Fig.~\ref{fig.2}, we set $k = 0$ and plot the real and imaginary parts of the lowest QNM frequencies normalized by $2 \pi T$ as a function of parameter $m/T$ for $\Delta = 2, 3$ and~$4$. One can clearly see that the frequencies do not change significantly (\ie by more than a factor of 2) as a function of $m/T$. This results in universal thermalization time set by the inverse of the temperature. Note that the~QNM frequencies are not, in general,  monotonic functions of $m/T$. For an explanation of why all the curves in Fig.~\ref{fig.2} are saturate in the low temperature regime, see the supplemental material. 

In Fig.~\ref{fig.k} we set $m/T = 4.8$ where (approximately) the biggest deviations from conformality are expected and plot $\Delta = 2, 3$ and $4$ QNM frequencies as a function of momentum normalized to temperature. We find not only that deviations from the ${\cal N} = 4$ SYM result are small, but also that up to an overall normalization, the momentum dependence is almost the same in all three cases.

\vspace{5 pt}
\emph{Thermalization time for physical ${\cal N}=2^{*}$ gauge theory 
scalar operators.--} We would like to extend analysis of the previous section to physical operators of the 
${\cal N}=2^*$ gauge theory. In \cite{Bobev:2013cja}, Bobev, Elvang, Freedman and Pufu (BEFP) showed that agreement between gravity calculations and exact QFT results in the ${\cal N}=2^*$ gauge theory,  obtained using localization techniques~\cite{Buchel:2013id}, requires 
embedding of the PW action \eqref{eq.Lmatter} into a larger consistent truncation  of five-dimensional ${\cal N}=8$ gauged supergravity. The matter part of the action \eqref{eq.Lgrav} then becomes
\begin{equation}
- {\cal L}_{matter} = - {\cal L}_{BEFP}=
12\frac{(\del\eta)^2}{\eta^2}+
\frac{4(\del{\vec X})^2}{(1-\vec{X}^2)^2}+
\calv
\label{befp}
\end{equation}
where the potential $\calv$ is 
\begin{equation}
\hspace{-3 pt}\calv=-\frac{4}{L^2}\left[\frac{1}{\eta^4}+2\eta^2\ \frac{1+\vec{X}^2}{1-\vec{X}^2}
-\eta^8\ \frac{(X_1)^2+(X_2)^2}{(1-\vec{X}^2)^2}
\right]\,.
 \label{pbefp}
\end{equation}
The vector $\vec{X}=\left(X_1,X_2,X_3,X_4,X_5\right)$ combines five of the scalars and $\eta$ is the sixth. The symmetry of the action reflects the symmetries of the dual gauge theory~\cite{Bobev:2013cja}: the two scalars $(X_1,X_2)$ form a doublet under the $U(1)_R$, 
while $(X_3,X_4,X_5)$ form a triplet under $SU(2)_V$ and $\eta$ is a singlet.  
The PW action \eqref{eq.Lmatter} from \eqref{befp} with 
\begin{equation}
e^\alpha= \eta\,,\ \ X_{1}  = \tanh{\chi}\,,\ \ X_2=X_3=X_4=X_5=0 \,.
\label{truncate}
\end{equation}
 \begin{widetext}
The effective action describing fluctuations of PW backgrounds within BEFP is obtained 
 linearizing \eqref{befp} in $X_2,X_3,X_4,X_5$ scalar fields:
 \begin{eqnarray}
 &&\dd {\cal L}\equiv  -\call_{BEFP}+\call_{PW}+\calo(X_i^4)\equiv  \dd\call_2+\dd\call_V\,,\nonumber\\
&&\dd{\cal L}_2=-(1+c)^2 (\del X_2)^2-\frac {1+c}{L^2}\left((c^2+c) \rho_6^{4/3}-4 (1+c) \rho_6^{1/3}
 +\frac{4(\del c)^2}{c^2-1}\right) (X_2)^2\,,\nonumber\\ 
&&\dd\call_V=-(1+c)^2 (\del \vec{X}_V)^2-\frac {1+c}{L^2}\left((c^2-1) \rho_6^{4/3}-4 (1+c) \rho_6^{1/3}
 +\frac{4(\del c)^2}{c^2-1}\right) (\vec{X}_V)^2\,, 
 \label{linear} 
 \end{eqnarray}
 \end{widetext}
where $\vec{X}_V=(X_3,X_4,X_5)$ and $\rho_6=e^{6\alpha}$, $c=\cosh 2\chi$. 
Since $\dd {\cal L} $ is $SU(2)_V$ invariant, it is enough to 
consider only one component of $\vec{X}_V$, say $X_3$. 
The scalars $X_2$ and $X_3$ correspond to operators $\calo_3$ and $\calo_2$ with dimensions $\Delta=3$ and 2 
of ${\cal N}=2^*$ gauge theory. Following \cite{Balasubramanian:2013esa}, we compute the lowest QNM
frequencies of these operators. The results in Fig.~\ref{fig.3} show once again that  the QNM frequencies have a very mild  dependence on $m/T$.

The matter fluctuations $\delta \alpha$ and $\delta \chi$ represent operators with $\Delta=2$ and $3$ in the ${\cal N}=2^*$ gauge theory. The results for the lowest QNM frequencies of $\delta \alpha$ and $\delta \chi$ are collected in Fig.~\ref{fig.4}. As in the previous examples, the frequencies exhibit a very mild dependence on $m/T$.

\vspace{5 pt}

\emph{Thermalization time for the energy-momentum tensor.--} Quasinormal modes of the energy-momentum tensor in the boundary theory are dual to metric perturbations in the bulk. Different polarizations 
decompose into decoupled sets of various helicities with respect to the propagation 
direction in the boundary \cite{Kovtun:2005ev}. The simplest helicity-2 fluctuations 
are always equivalent to a minimally coupled massless scalar \cite{Buchel:2004qq} and its momentum-dependence (at $m/T = 4.8$) is given by $\Delta = 4$ curve in Fig.~\ref{fig.k}. In the zero-momentum limit, the helicity distinction vanishes implying that generic metric perturbations obey the equation of motion for a minimally coupled massless scalar. Hence in this case, the QNM are the same as the scalar operator with $\Delta=4$ plotted in Fig.~\ref{fig.2}. Finally, in Fig.~\ref{fig.4}, we plot zero-momentum QNM associated with the perturbations of the background scalars $\alpha$ and $\chi$. Again, we find rather mild dependence on $m/T$.

\vspace{5 pt}

\emph{Comparison to conformal plasma at nonzero chemical potential.--} From an operational point of view, deviations from conformality in ${\cal N} = 2^{*}$ gauge theory amount to the presence of a dimensionless parameter $m/T$ characterizing the equilibrium configuration. In order to confirm the robustness of our main result, we consider a strongly coupled conformal QGP at nonzero chemical potential $\mu$. In this case, the dimensionless parameter  is $\mu/T$. The dual solution is the anti-de Sitter Reissner-Nordstrom black hole (AdS-RN). We compute the lowest zero-momentum quasinormal modes of phenomenological operators (with $\Delta = 2$, $3$ and $4$) dual to probe (neutral) scalar fields in the AdS-RN background.

Again, as shown in Fig. \ref{fig.RN}, the frequencies depend mildly on $\mu/T$ unless the chemical potential (at fixed temperature) gets big enough to see the effects of the critical behaviour associated in the dual description with the (near-)extremal AdS-RN. In this case, the dominant QNM has a purely imaginary frequency. We explain in the supplemental material why we do not expect similar transition to occur in ${\cal N} = 2^{*}$ gauge theory and refer the reader to refs \cite{Edalati:2010hk} and \cite{Bhaseen:2012gg}, which discuss these purely imaginary QNM in detail.

\vspace{5 pt}

\emph{Note added:} While this letter was being finalized we learnt about the upcoming results of Ref.~\cite{JanikSpalinski}, which presents the computation of QNM in a bottom-up holographic QCD model. The results of \cite{JanikSpalinski} are in line with the point of view presented in this letter, i.e. equilibration rates in strongly coupled nonconformal plasma are not very different from those of ${\cal N} = 4$ SYM. 

\vspace{5 pt}

\begin{acknowledgements}
We would like to thank A. Donos, J. Fuini, U. Gursoy, R. Janik, J. Jottar, M. Kaminski, L. Lehner, H. Soltanpanahi, M. Spalinski, W. van der Schee and I. Yavin for useful discussions, correspondence and comments on the draft. We would like to thank the authors of articles \cite{FuiniYaffe} and \cite{JanikSpalinski} for generously sharing their drafts prior to publication. This work was partly supported
by the National Science Centre grant 2012/07/B/ST2/03794. Research at Perimeter Institute is supported by the Government of Canada through
Industry Canada and by the Province of Ontario through the Ministry of
Research \& Innovation. AB and RCM acknowledge support from NSERC Discovery grants. RCM also acknowledges funding from the Canadian Institute for Advanced Research.

\end{acknowledgements}

\vspace{10 pt}

\begin{figure}
\includegraphics[width=0.47\textwidth]{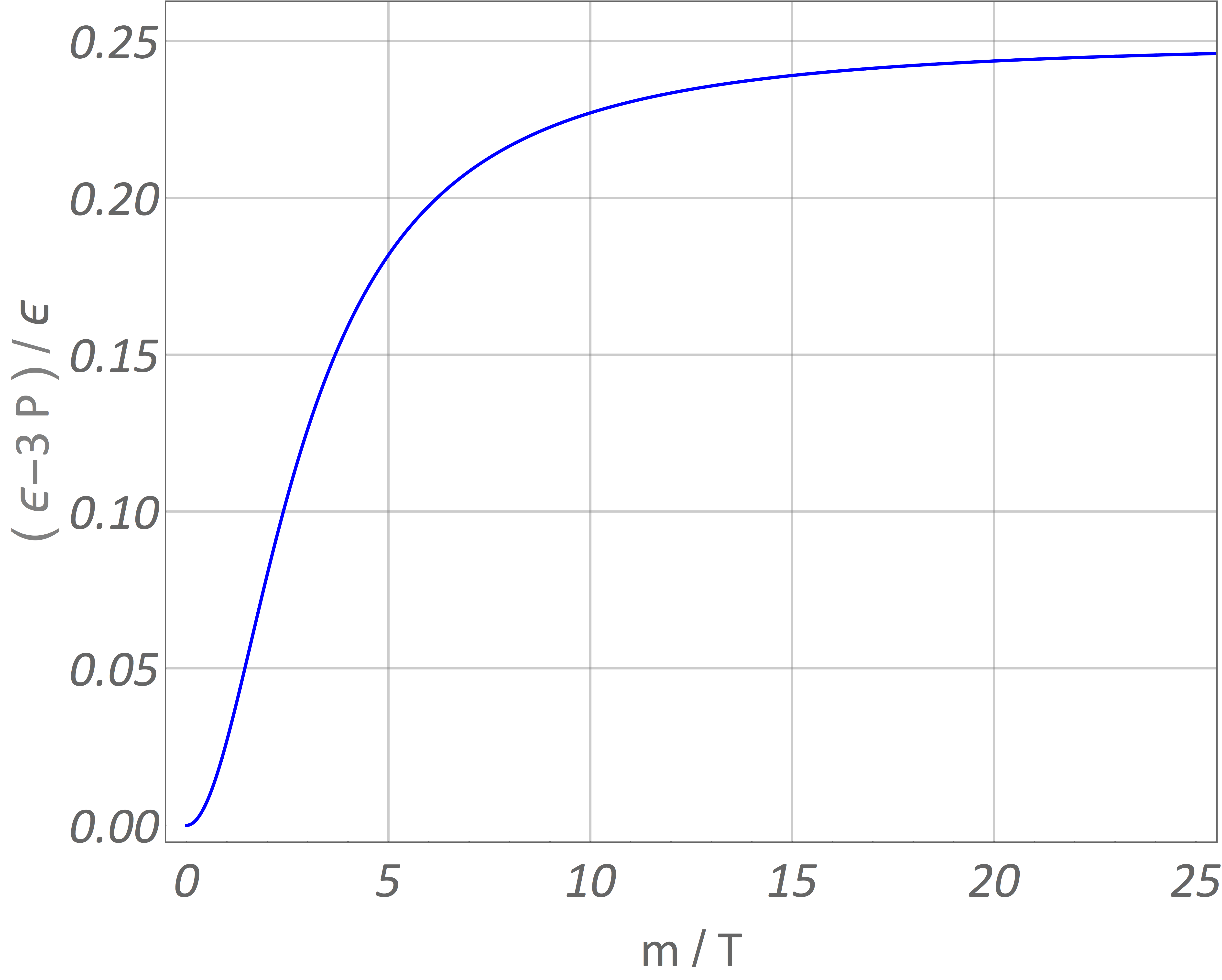}
\caption{Trace of the energy-momentum tensor normalized to the energy density as a function of $m/T$ (see also Fig.~\ref{fig.1}). This quantity saturates at low temperatures to the value consistent with the thermodynamics of a five-dimensional CFT.}
\label{fig.7}
\end{figure}

\emph{Supplemental material.--} Ref. \cite{HoyosBadajoz:2010td}, following \cite{Buchel:2007mf}, observed that for large $m/T$ the thermodynamics and the lowest transport coefficients of the four-dimensional \mbox{${\cal N} = 2^{*}$} gauge theory are inherited from a five-dimensional conformal field theory (CFT) compactified on a circle. This is apparent in Fig. \ref{fig.7}, where the ratio of the trace of the thermal stress tensor to the energy density approaches 0.25 for large $m/T$. In a five-dimensional CFT, the energy density and pressure are related by $\epsilon_{5} = 4 P_{5}$, which indeed gives $\left( \epsilon_{5} -3 \, P_{5} \right) / \epsilon_{5} = 1/4$.

Below we will demonstrate that this near-horizon AdS region dominates also the behaviour of the QNM in the regime of large $m/T$. This constitutes a highly nontrivial test for our numerics and demonstrates that we did not miss any exotic branches of QNM's. To set the background, let us thus briefly remark on the relation between the five-dimensional bulk geometry in the low temperature regime ($m/T \gg 1$) and an auxiliary six-dimensional dual of a finite-temperature CFT \cite{HoyosBadajoz:2010td}. At low enough temperature the field $\chi$ attains a large value in the infra-red (IR) region of the five-dimensional geometry 
\eq
\label{eq.chialphaIR}
e^{2\chi}\simeq 2u\,,\quad e^{6\alpha}\simeq \frac{2}{3u}\,,\quad e^A\simeq \left(\frac{2}{3u^4}\right)^{1/3}(mL),
\eqx
where we introduced a new radial coordinate $u$. Defining
\eq
\label{eq.phi12}
\phi_{1} = \frac{1}{2} \left( 3\alpha + \chi \right) \quad \mathrm{and} \quad \phi_{2} = \frac{1}{2} \left(\alpha - \chi \right)
\eqx
one can lift the bulk geometry \reef{eq.metric} to six dimensions with
\eq
\label{eq.lift}
ds_{6}^{2} = e^{-2 \phi_{2}} ds_{5}^{2} + 4m^2L^2\ e^{6 \phi_{2}} dx_{6}^{2}\,.
\eqx
In particular, the zero-temperature geometry recovered in the IR using \eqref{eq.lift} is a six-dimensional AdS spacetime
\eq
\label{irgeom}
\hspace{-4 pt} ds_{5,1}^2\simeq \frac{3^{3/2} L^2}{2u^2}\left[du^2+ \frac{4m^{2}}{9}\left( \eta_{\mu\nu}dx^\mu dx^\nu+ dx_6^2\right)\,\right]
\eqx
where the six-dimensional curvature scale is $L_6/L={3^{3/4}}/{2^{1/2}}$.

Let us now reconsider the QNM of phenomenological operators from the point of view of the six-dimensional geometry in the IR. Uplifting the scalar field action \eqref{eq.Sscalar} to six dimensions using \eqref{eq.lift} gives
\eq
\label{eq.Sscalaruplift}
S \sim \int \mathrm{d}^{6} x \sqrt{-\tilde{g}} \left( (\tilde{\partial} \psi)^{2} + e^{2 \phi_{2}} M^{2} \psi^{2}\right), 
\eqx
where tilde denotes quantities defined using the six-dimensional metric \eqref{eq.lift}. As in the relevant regime $\phi_{2}$ approaches $-\infty$, see eqs.~\eqref{eq.chialphaIR} and \eqref{eq.phi12}, the mass term in eq.~\eqref{eq.Sscalaruplift} becomes negligible and the resulting QNM frequencies must be that of a massless scalar field in a six-dimensional AdS-Schwarzschild black hole. Recalling the holographic dictionary for scalars in AdS$_6$/CFT$_5$, 
\eq
\label{eq.MfuncDelta}
M^{2} = \frac{1}{L^{2}} \Delta(\Delta - 5)\,,
\eqx
we see the massless scalar in the effective six-dimensional geometry is dual to a marginal operator in the dual five-dimensional CFT. 
This is in excellent agreement with our numerics, as fits to the large-$m/T$ behaviour of frequencies plotted in Fig. \ref{fig.2} are all consistent with the lowest QNM of a marginal operator in a five-dimensional holographic CFT. We performed similar analysis for the other fields which we considered and found analogous behaviour with the exception that the effective IR masses in uplifted actions need not always be zero. Let us also note that in Fig.~\ref{fig.3} and~\ref{fig.4} the frequencies saturate to their IR values much more slowly than in Fig.~\ref{fig.2} because of complicated couplings of the corresponding perturbations to the background fields. 

This result is interesting on its own right. To the best of our knowledge, it provides the first example in holography where the low temperature non-equilibrium response (\ie away from $\omega \rightarrow 0$) of strongly coupled matter is \emph{fully} determined by the properties of the IR geometry in the dual gravity description.

\bibliography{nonconf_biblio}{}

\end{document}